\documentclass[aps,prd,twocolumn,superscriptaddress,preprintnumbers,floatfix,nofootinbib]{revtex4-1}

\usepackage{graphicx}
\usepackage{dcolumn}
\usepackage{bm}
\usepackage{hyperref}
\usepackage{float}
\usepackage{color}

\newcommand{\be}{\begin{equation}}
\newcommand{\ee}{\end{equation}}
\newcommand{\bea}{\begin{eqnarray}}
\newcommand{\eea}{\end{eqnarray}}

\usepackage{mathrsfs}
\usepackage{amsmath, amsthm, amssymb}
\usepackage{color}
\usepackage{epstopdf}
\usepackage{amssymb}
\usepackage{verbatim}
\usepackage{hyperref}
\usepackage{enumerate}
\usepackage{float}
\usepackage[caption = false]{subfig}
\usepackage[normalem]{ulem}  
\usepackage{epsfig}

\newcommand{\apjl}{ApJLett}		
\newcommand{\mnras}{MNRAS}		

\newcommand{\physrep}{Phys.~Rep.}   

\begin{document}

\title{Gravitational Bose-Einstein Condensation of Vector/Hidden Photon Dark Matter}

\author{Jiajun Chen}
\thanks{Contributes equally to this work}
\email{chenjiajun@swu.edu.cn}
\affiliation{School of Physical Science and Technology, Southwest University, Chongqing 400715, China}

\author{Xiaolong Du}
\thanks{Contributes equally to this work}
\email{xdu@astro.ucla.edu}
\affiliation{Department of Physics and Astronomy, University of California, Los Angeles, CA 90095, USA}
\affiliation{Carnegie Observatories, 813 Santa Barbara Street, Pasadena, CA 91101, USA}

\author{Mingzhen Zhou}
\email{zhoumz@swu.edu.cn}
\affiliation{School of Physical Science and Technology, Southwest University, Chongqing 400715, China}

\author{Andrew Benson}
\email{abenson@carnegiescience.edu}
\affiliation{Carnegie Observatories, 813 Santa Barbara Street, Pasadena, CA 91101, USA}

\author{David J. E. Marsh}
\email{david.j.marsh@kcl.ac.uk}
\affiliation{Theoretical Particle Physics and Cosmology, King's College London, Strand, London, WC2R 2LS, United Kingdom}

\begin{abstract}
We study the gravitational Bose-Einstein condensation of a massive vector field in the kinetic regime and the non-relativistic limit using non-linear dynamical numerical methods. Gravitational condensation leads to the spontaneous formation of solitons. We measure the condensation time and growth rate, and compare to analytical models in analogy to the scalar case. We find that the condensation time of the vector field depends on the correlation between its different components. For fully correlated configurations, the condensation time is the same as that for a scalar field. On the other hand, uncorrelated or partially correlated configurations condense slower than the scalar case. As the vector soliton grows, it can acquire a net spin angular momentum even if the total spin angular momentum of the initial conditions is zero.
\end{abstract}

\maketitle

\section{Introduction}

The composition of dark matter (DM) is one of the most important unresolved problems in modern cosmology. Observations show dark matter makes up about $27\%$ of the total mass–energy in our Universe \cite{Aghanim:2018eyx}. One popular idea is that dark matter is composed of a light (sub-eV) scalar or pseudo-scalar field, the canonical example of which is the QCD axion or other axion-like particles~\cite{Dine:1982ah,Suarez:2013iw,Preskill:1982cy,Abbott:1982af,2015PhRvD..92j3513G,Widrow:1993qq,Uhlemann:2014npa}. We refer to such models in the following as ``scalar dark matter'' (SDM).
SDM can form solitons balanced by competing forces of gravity and gradient energy~\cite{1968PhRv..172.1331K,PhysRevD.42.384,1993PhRvL..71.3051K,Widrow:1993qq,2015MNRAS.451.2479M,Hui:2016ltb}. We also call these solitons ``boson stars''. Boson stars are known to form via two mechanisms. The first mechanism occurs during the initial stages of collapse of a halo, where coherence is inherited from the Jeans scale in the initial conditions~\cite{Schive:2014dra,Eggemeier:2019jsu}. The second mechanism occurs due to gravitational Bose-Einstein condensation in the kinetic regime, whereby boson stars form spontaneously~\cite{Levkov:2018kau}. See also Ref.~\cite{Chavanis:2020upb} and references therein for a more detailed discussion on formation mechanisms for boson stars.

It is also possible for spin-1 particles to obtain small masses and thus constitute light bosonic DM~\cite{Arias:2012az}. Such models are commonly known as ``hidden/dark photon'' DM, and in the following we refer to them generically as ``vector DM'' (VDM). VDM has become popular in recent years since viable production mechanisms, similar to vacuum realignment of scalars, have been proposed~\cite{Graham:2015rva}. Like light axion-like particles~\cite{axiverse,Mehta:2021pwf}, light hidden photons also arise in string compactifications~\cite{Goodsell:2009xc}. Furthermore, direct detection of VDM has developed rapidly in recent years (see Ref.~\cite{Caputo:2021eaa} and references therein). Indeed, direct VDM searches can often be performed as trials for axion searches, since dark photon kinetic mixing operates similarly to axion-photon conversion without the need for an external magnetic field. Thus the proliferation of axion direct searches naturally leads to a proliferation in VDM searches during the early years of technology development. 

VDM, like SDM, can also form solitons, known sometimes as ``Proca stars''~\cite{Brito:2015pxa,Liebling:2012fv}. A real scalar field has a single degree of freedom, while VDM has three degrees of freedom. Thus, VDM solitons can be polarised~\cite{Jain:2021pnk}. VDM solitons have recently been observed to form in cosmological simulations~\cite{Gorghetto:2022sue}, using the first mechanism outlined above, i.e. initial coherence. The decay of VDM solitons via parametric resonance has been discussed in \cite{Amin:2023imi}. In the non-relativistic limit, the dynamics of VDM can be simulated by a multicomponent SDM system \cite{Zhang:2021xxa,Gorghetto:2022sue,Adshead:2021kvl,Jain:2021pnk,Gosenca:2023yjc,Glennon:2023jsp}. In the following we demonstrate for the first time gravitational condensation of VDM boson stars using methods developed by us previously for SDM~\cite{Chen:2020cef,Chen:2021oot}.

We study the condensation of VDM from different initial conditions with and without correlation between different components of the vector field. We find the following:
\begin{itemize}
\item If different components of VDM are fully correlated, e.g. extremely polarized initial conditions, the condensation of VDM happens in the same way as the SDM;

\item If different components of VDM are uncorrelated, i.e. each component is an independent Gaussian random field, the condensation time scale of VDM is longer than that of SDM with the same mean density. The condensation time is determined by the component with the highest mean density.

\item If different components of VDM are partially correlated, condensation happens on time scales between the previous two cases. The condensation time can be estimated from the eigenvalues of the covariance matrix.
\end{itemize}

This paper is organised as follows. In Sec.~\ref{sec:EOM} we derive the equation of motion for VDM in the non-relativistic limit and show how we generate the initial conditions. In Sec.~\ref{sec:symmetry}, we revisit a few symmetries and conservation laws for the Schr\"{o}dinger-Poisson vector system. In Sec.~\ref{sec:condense}, we discuss the condensation of VDM and the formation of Proca stars from homogeneous and isotropic initial conditions with and without initial correlation between different components of the vector field. We present our simulation results for the condensation time scales for VDM in Sec.~\ref{sec:relax_time} and conclude in Sec.~\ref{sec:conclusion}.

\section{Equations of Motion and Initial Conditions}\label{sec:EOM}

A real vector field, $A_\mu$, with mass $m$, minimally coupled to gravity, has the following action:
\begin{equation}
S =\int d^4 x \sqrt{-g}\left( -\frac{1}{4}F_{\mu\nu}F^{\mu\nu}-\frac{1}{2}m^2 A_\mu A^\mu\right)\, ,
\label{eq:action}
\end{equation}
where $F_{\mu\nu}=\partial_\mu A_\nu-\partial_\nu A_\mu$ is the field strength tensor, and indices are raised and lowered by the metric $g_{\mu\nu}$ with determinant $g$. The equation of motion is:
\begin{equation}
\nabla_\mu F^{\mu\nu}= m^2 A^\nu\, ,
\label{eq:proca_eq}
\end{equation}
known as the Proca equation, which implies the Lorentz condition, $\nabla_\mu A^\mu=0$, as a constraint, thus reducing the number of degrees of freedom of $A_\mu$ to three dynamical fields (see e.g. Refs.~\cite{Brito:2015pxa,Gorghetto:2022sue}). We consider the non-relativistic limit, where these degrees of freedom are written as:
\begin{equation}
A_i = \frac{1}{\sqrt{2m}}\left( \psi_i e^{imt}+\psi_i^* e^{-imt}\right)\, .
\label{eq:A_i}
\end{equation}

In this limit, in exact analogy to a real scalar field (see e.g. Ref.~\cite{Marsh:2015xka}), for small perturbations of the metric about Minkowski space by the Newtonian potential $V$, and at lowest order in the non-relativistic limit the Einstein-Proca equations reduce to three copies of the Schr\"{o}dinger-Poisson (SP) equations for each complex field $\psi_i$:
\begin{eqnarray}
i\frac{\partial}{\partial{t}} \bm{\Psi}&=&-\frac{1}{2m}\nabla^2\bm{\Psi} + m V\bm{\Psi},
\label{eq:SP1}
\\
\nabla^2{V}&=&4 \pi G m\left(\bm{\Psi}^{\dag}\bm{\Psi}-n\right),
\label{eq:SP2}
\end{eqnarray}
where $\bm{\Psi}=(\psi_1,\psi_2,\psi_3)^T$ is the wave function vector, $n=n_1+n_2+n_3$ is the mean total number density, $G$ is Newton's gravitational constant, and natural units, $\hbar=c=1$, are used.

Introducing the dimensionless quantities
\begin{eqnarray}
x=\widetilde{x}/(m v_0),\quad t=\widetilde{t}/(mv_0^2),\quad V=\widetilde{V} v_0^2,\nonumber\\
\psi_i=\widetilde{\psi_i}v_0^2\sqrt{m/(4 \pi G)},
\label{eq:dim}
\end{eqnarray}
where $v_0$ is a reference velocity (e.g. the characteristic velocity of the initial state), we obtain the dimensionless equations that we will solve numerically
\begin{eqnarray}
i\frac{\partial}{\partial{\widetilde{t}}}\widetilde{\bm{\Psi}}&=&-\frac{1}{2}\widetilde{\nabla}^2\widetilde{\bm{\Psi}} + \widetilde{V}\widetilde{\bm{\Psi}},
\label{eq:SP1_dim}
\\
\widetilde{\nabla}^2{\widetilde{V}}&=&\widetilde{\bm{\Psi}}^{\dag}\widetilde{\bm{\Psi}}-{\widetilde{n}}.
\label{eq:SP2_dim}
\end{eqnarray}

Similar to previous studies on the condensation of SDM~\cite{Levkov:2018kau,Chen:2020cef,Chen:2021oot}, we generate homogeneous and isotropic initial conditions in a periodic box of size $L$. Each component of the vector field is assumed to be a Gaussian random field with a momentum distribution of $|\psi_i(\bm{p})|^2 \propto \delta(|\bm{p}|-m v_0)$. Other choices of momentum distribution such as a Gaussian or Heaviside step function have been studied for SDM in previous literature. Ref.~\cite{Levkov:2018kau} shows that different choices of initial momentum distribution only lead to a minor change to the overall coefficient in the analytic formula of condensation time. Thus we limit our study to the Dirac-$\delta$ initial distribution. Specifically, given the momentum distribution we perform an inverse Fourier transform on $\psi_i(\bm{p}) e^{i S}$ with $S$ a random phase. The wave function is normalized to give a specified total number of non-relativistic bosons for the $i-$th component in the box, $N_i\equiv n_i L^3$. The initial wave function is then evolved by solving the SP equations numerically using a fourth-order time-splitting pseudospectral method \cite{Du:2018qor,Mocz:2017wlg}.

\section{Symmetry and Conservation Law}\label{sec:symmetry}
Before showing our simulation results, it is helpful to revisit a few important symmetries and conservation laws of the VDM system in the nonrelativistic limit. Firstly, each component of $\bm{\Psi}$ has a global $U(1)$ symmetry, $\psi_i \rightarrow e^{i\phi}\psi_i$, which leads to the conversation of particle number for individual component
\begin{equation}
N_i=\int d^3 x\,\psi_i^* \psi_i.
\label{eq:conser_N}
\end{equation}
Secondly, the SP Eqs.~(\ref{eq:SP1}) and (\ref{eq:SP2}) are invariant under $3$D rotation
\begin{equation}
\bm{\Psi} \rightarrow R \bm{\Psi},
\label{eq:rot}
\end{equation}
where $R$ is the rotation matrix. The $SO(3)$ symmetry leads to the conservation of orbital angular momentum for individual component, $\bm{L_i}$, and intrinsic spin angular momentum, $\bm{S}$:
\begin{eqnarray}
\bm{L}_i &=& \int d^3 x\, \left[\bm{r} \times \frac{1}{2 i}\left(\bm{\psi_i}^{*}\nabla \bm{\psi_i}-\bm{\psi_i}\nabla\bm{\psi_i}^* \right)\right],
\label{eq:conser_L}\\
\bm{S} &=& \int d^3 x\, \left(i \bm{\Psi} \times \bm{\Psi}^{*}\right).
\label{eq:conser_S}
\end{eqnarray}
Here the intrinsic spin angular momentum is an additional property of VDM which does not exist for SDM.
Finally, we have energy conservation
\begin{equation}
E=\int d^3 x \, \left[ \sum_i \frac{1}{2 m} \left|\nabla\psi_i\right|^2-\frac{1}{2} m \bm{\Psi}^{\dag} \bm{\Psi} V\right].
\label{eq:conser_E}
\end{equation}

Another important feature of the SP equations (\ref{eq:SP1}) and (\ref{eq:SP2}) is that if $\bm{\Psi}$ is one solution, we can do a unitary transformation
\begin{equation}
\bm{\Psi}' = U \bm{\Psi},
\label{eq:unitory_trans}
\end{equation}
where $U$ is a unitary matrix satisfying $U^{\dag}U=1$, the new vector field $\bm{\Psi}'$ is also a solution to the SP equations and the dynamical evolution of the system is unaffected. For example, the evolution of total density distribution will be identical before and after the transformation because $\bm{\Psi}'^{\dag}\bm{\Psi}'=\bm{\Psi}^{\dag}\bm{\Psi}$. We will come back to this point later and show that this important property will help us understand the similarity and difference between the condensation time scales for VDM and SDM. However, it should be noted that in general the spin of the new field $\bm{S}'\neq U \bm{S}$, so the spin density will be different after the transformation (see also the discussion in Ref.~\cite{Amin:2022pzv}).

In the current work, we neglect the possible self-interactions which may partially break some of the symmetries mentioned above \cite{Jain:2022agt,Jain:2023qty}. For example, the particle number and angular momentum of individual components will no longer be conserved, instead we expect the total particle number and total angular momentum are conserved. Furthermore, the dynamical evolution of the system will also be affected if the self-interactions contain a spin-spin interaction term. We leave the study of condensation of VDM with self-interactions to future work.

\section{Condensation of VDM by gravitational interaction}\label{sec:condense}

The condensation of SDM through gravitational relaxation \cite{Levkov:2018kau,Chan:2022bkz} and/or self-interaction \cite{Chen:2020cef,Kirkpatrick:2020fwd,Chen:2021oot,Kirkpatrick:2021wwz} has been studied extensively. Neglecting the self-interaction, which is usually a sub-dominant effect, the condensation time scale due to gravity is found to be in good agreement with the analytic formula \cite{Levkov:2018kau}:
\begin{equation}
\tau_{\rm gravity}^S =\frac{4b\sqrt{2}}{\sigma_{\rm gravity}v n f},
\label{eq:condensation_time_scalar}
\end{equation}
where $\sigma_{\rm gravity}=8\pi(m G)^2\lambda/v^4$ is the the transport Rutherford cross section of gravitational scattering with $\Lambda=\ln(m v L)$ the Coulomb logarithm, $f=6\pi^2 n/(m v)^3$ is the phase-space density, and $b$ is an $O(1)$ coefficient. For ultralight bosonic particles, $f \gg 1$, thus enhancing the relaxation rate due to gravitational scattering, which is called ``Bose stimulation". Substituting $\sigma_{\rm gravity}$ and $f$ into Eq.~(\ref{eq:condensation_time_scalar}), the condensation time is given by
\begin{equation}
\tau_{\rm gravity}^S = \frac{b\sqrt{2}}{12\pi^3}\frac{m v^6}{G^2 n^2 \Lambda}.
\label{eq:tau_S}
\end{equation}

The condensation of SDM particles leads to the formation of Bose stars whose radial density profile is well described by the soliton solution~\cite{Schive:2014dra,Schive:2014hza,2015MNRAS.451.2479M,Levkov:2018kau}
\begin{equation}
\rho_{\rm soliton}(r)=\rho_0\left[1+9.1\times 10^{-2}\left(\frac{r}{r_{\rm c}}\right)^2\right]^{-8},
\label{eq:soliton_profile}
\end{equation}
where $\rho_0$ is the central density, $r_c$ is the core radius at which the density drops to one half of the central density. Due to the scaling symmetry of the SP equations (e.g. Ref.~\cite{Schive:2014dra}), $\rho_0$ and $r_c$ are not independent of each other:
\begin{equation}
\rho_0 = 1.9\times 10^{7}\left(\frac{m}{10^{-22}{\rm eV}}\right)^{-2}\left(\frac{r_c}{{\rm kpc}}\right)^{-4}M_{\odot}{\rm kpc}^{-3}.
\label{eq:rho_0}
\end{equation}

Now let us consider the VDM model. In the remainder of this paper, we will work in the Cartesian basis $\{\hat{e}_x,\hat{e}_y,\hat{e}_z\}$ for simulation purposes, unless otherwise specified. \footnote{Equivalently, one can also choose to work in the spin basis $\{\epsilon_z^{(1)}, \epsilon_z^{(-1)}, \epsilon_z^{(0)}\}$ given by Eq. (44) in \cite{Jain:2021pnk}.}  For VDM, we have extra degrees of freedom to set the initial conditions for each component. In the following subsections, we will discuss different choices of initial conditions based on the correlation between different components and compare the results with SDM.

\subsection{Fully correlated initial conditions}
The simplest choice is to have only one nonzero component for the vector field, e.g. $\bm{\Psi} = (0, 0, \psi_z)^T$. In such a case, the SP equations reduce to the ones for SDM, so vector particles condense in the same way as SDM. Furthermore, as we have argued in Sec.~\ref{sec:symmetry}, if we perform a unitary transformation, $\bm{\Psi}'=U\bm{\Psi}$, the condensation time due to gravity does not change. For example, the extremely polarized initial states can be written as \cite{Jain:2021pnk}
\begin{equation}
\bm{\Psi}^{(\pm 1)}=
\frac{1}{\sqrt{2}}
\begin{pmatrix}
        1 \\
    \pm i \\
        0
\end{pmatrix}
\psi,
\quad
\bm{\Psi}^{(0)}=
\begin{pmatrix}
        0 \\
        0 \\
        1
\end{pmatrix}
\psi,
\label{eq:polarized}
\end{equation}
where $\psi$ is a Gaussian random field, and $\bm{\Psi}^{(\pm 1)}$, $\bm{\Psi}^{(0)}$ correspond to circular and linear polarized states, respectively. The circular polarized states can be transformed to the linear polarized state by a unitary transformation with
\begin{equation}
U^{\pm}=
\begin{pmatrix}
   \frac{1}{2}        & \pm\frac{i}{2}         &    \frac{1}{\sqrt{2}} \\
\mp\frac{i}{2}        &    \frac{1}{2}         & \pm\frac{i}{\sqrt{2}} \\
   \frac{1}{\sqrt{2}} & \mp\frac{i}{\sqrt{2}}  &  0
\end{pmatrix}
.
\label{eq:unitary_matrix}
\end{equation}
Thus as the linear polarized case, $\bm{\Psi}^{(0)}$, with only one nonzero component, the condensation time for $\bm{\Psi}^{(\pm 1)}$ is also the same as SDM. \footnote{Note that the intrinsic spin angular moment has changed and $S'\neq U^{\pm}S$ as discussed in Sec.~\ref{sec:symmetry}. But since we focus on the density evolution, it does not affect our conclusion on the condensation time scale.} In general, the conclusion can be extended to any initial conditions that can be unitarily transformed to $\bm{\Psi}^{(0)}$. Essentially, for all these initial conditions, different components of the vector field are fully correlated and can be represented by a single scalar field up to a global complex constant.

\subsection{Uncorrelated initial conditions}
In the previous subsection, we have shown that fully correlated initial conditions for VDM condense on the same time scale as SDM. How does it change if different components of VDM are initially independent of each other (uncorrelated)? To answer this question, we can follow the same calculations in \cite{Levkov:2018kau,Kirkpatrick:2020fwd} that consider the kinetic equation for the homogeneous isotropic ensemble with long-range gravitational interaction. Basically, the bosons condense on the kinetic relaxation time scale given by Eq. (\ref{eq:condensation_time_scalar}). Since we have not included self-interactions or spin-dependent interactions, we expect that the cross section remains the same. However, as we mentioned earlier, the vector field has extra degrees of freedom. As is shown in \cite{Jain:2021pnk}, VDM allows solitonic solutions that have the same density profile [see Eq.~(\ref{eq:soliton_profile})] as SDM. For VDM, however, there exist three extremely polarized states with the same energy. In the Cartesian basis, this is equivalent to saying that there are three possible solitonic solutions which are orthogonal to each other:
\begin{equation}
\bm{\Psi}_{\rm soliton} = \sqrt{\rho_{\rm soliton}(r)} e^{-i E_s t} \hat{e}_i, \quad i=x,y,z.
\label{eq:sol_VDM}
\end{equation}

When VDM particles condense, they can condense to either of above solitonic solutions depending on the initial conditions. So we propose that when computing the ``Bose stimulation" effect, we should compute the phase-space density for each component separately
\begin{equation}
f_i=\frac{6\pi^2 n_i}{(m v)^3}.
\label{eq:f_i}
\end{equation}
The condensation time of the $i$-th component is then given by
\begin{equation}
\tau_{\rm gravity,i} = \frac{b\sqrt{2}}{12\pi^3}\frac{m v^6}{G^2 n_i n \Lambda},
\label{eq:tau_i}
\end{equation}
where $n=n_x+n_y+n_z$ is the mean total number density. Note we still have a term in the denominator that is proportional to $n$ instead of $n_i$ which comes from the scattering rate.

One prediction from Eq.~(\ref{eq:tau_i}) is that if the mass is equally distributed among different components such that $n_x=n_y=n_z=n/3$, the condensation time of VDM will be $3$ times longer than for SDM with the same mean total number density. We check our prediction against numerical simulations.

\begin{figure}[htbp]
\includegraphics[width=\columnwidth]{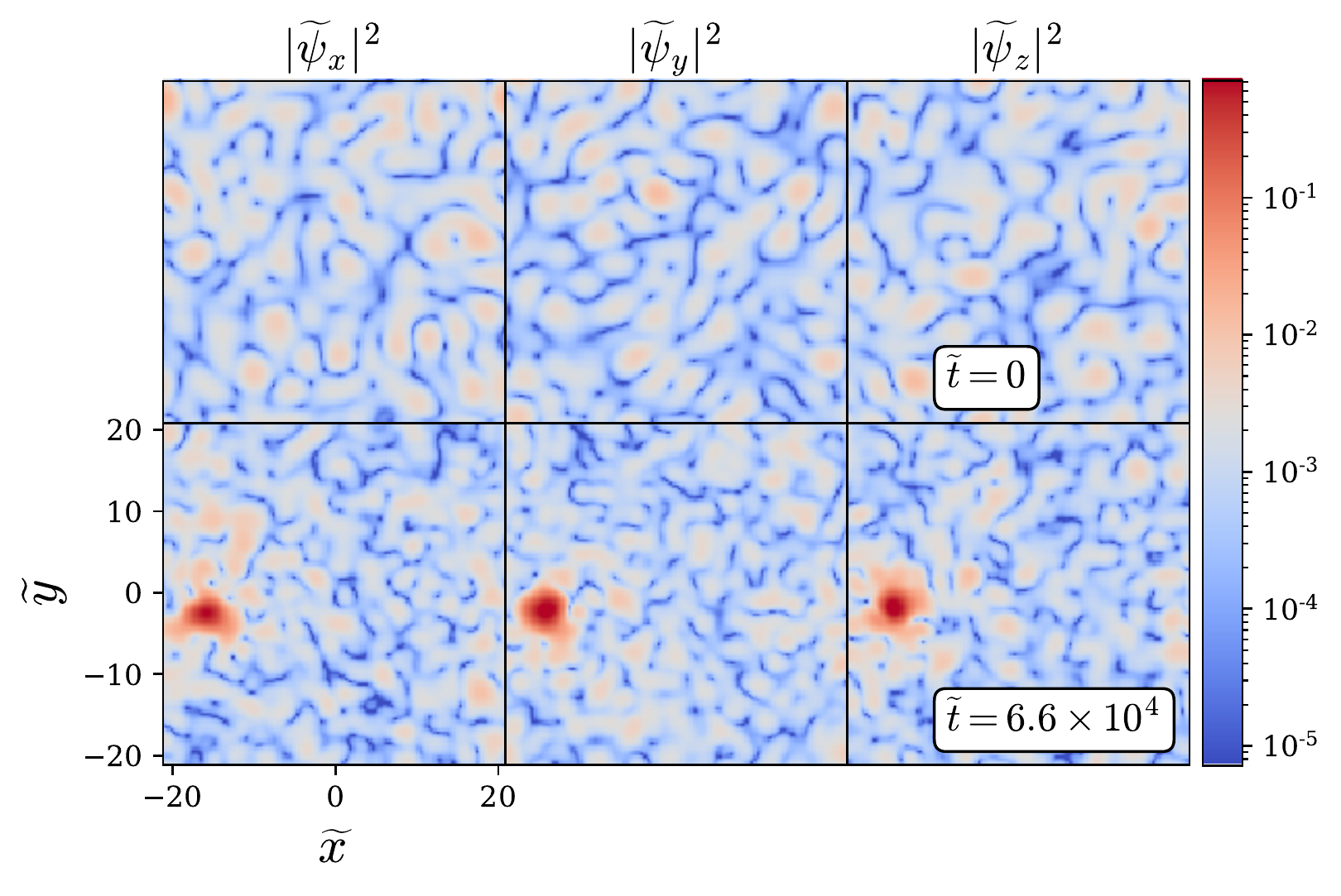}
\caption{Formation of a Proca star from homogeneous and isotropic initial conditions with $\widetilde{N}_1 = \widetilde{N}_2 = \widetilde{N}_3 = 83.8$, $\widetilde{L}=42$. Upper panel: slice density field for each component of the vector field at the initial time. Lower panel: slice density field through the center of the Proca star at $\widetilde{t}=2\,\widetilde{\tau}_{\rm gravity}$.}
\label{fig:slice_N_84_L_42}
\end{figure}

Firstly, we generate initial conditions with equal number of particles for individual components. Each component is an independent Gaussian random field with a $\delta$ momentum distribution as described in Sec. \ref{sec:EOM}. One example of our simulations with $\widetilde{L}=42$  and $\widetilde{N}_1 = \widetilde{N}_2 = \widetilde{N}_3 = 83.8$ is shown in Fig.~\ref{fig:slice_N_84_L_42}. As can be seen, a compact spherical object, a Proca star (lower panels), forms from homogeneous and isotropic initial conditions (upper panels). 

\begin{figure*}[t]
\includegraphics[width=\columnwidth]{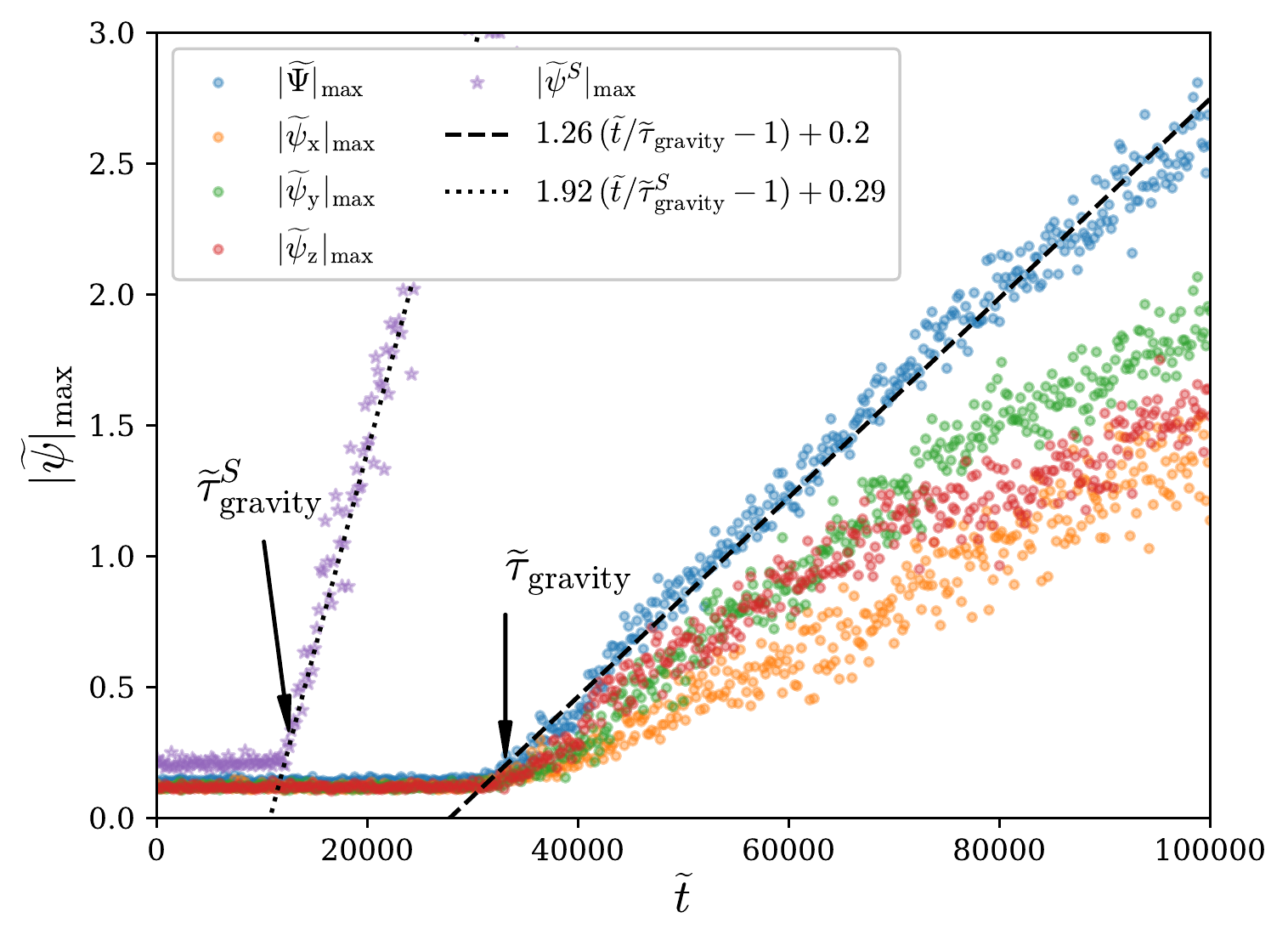}
\includegraphics[width=0.95\columnwidth]{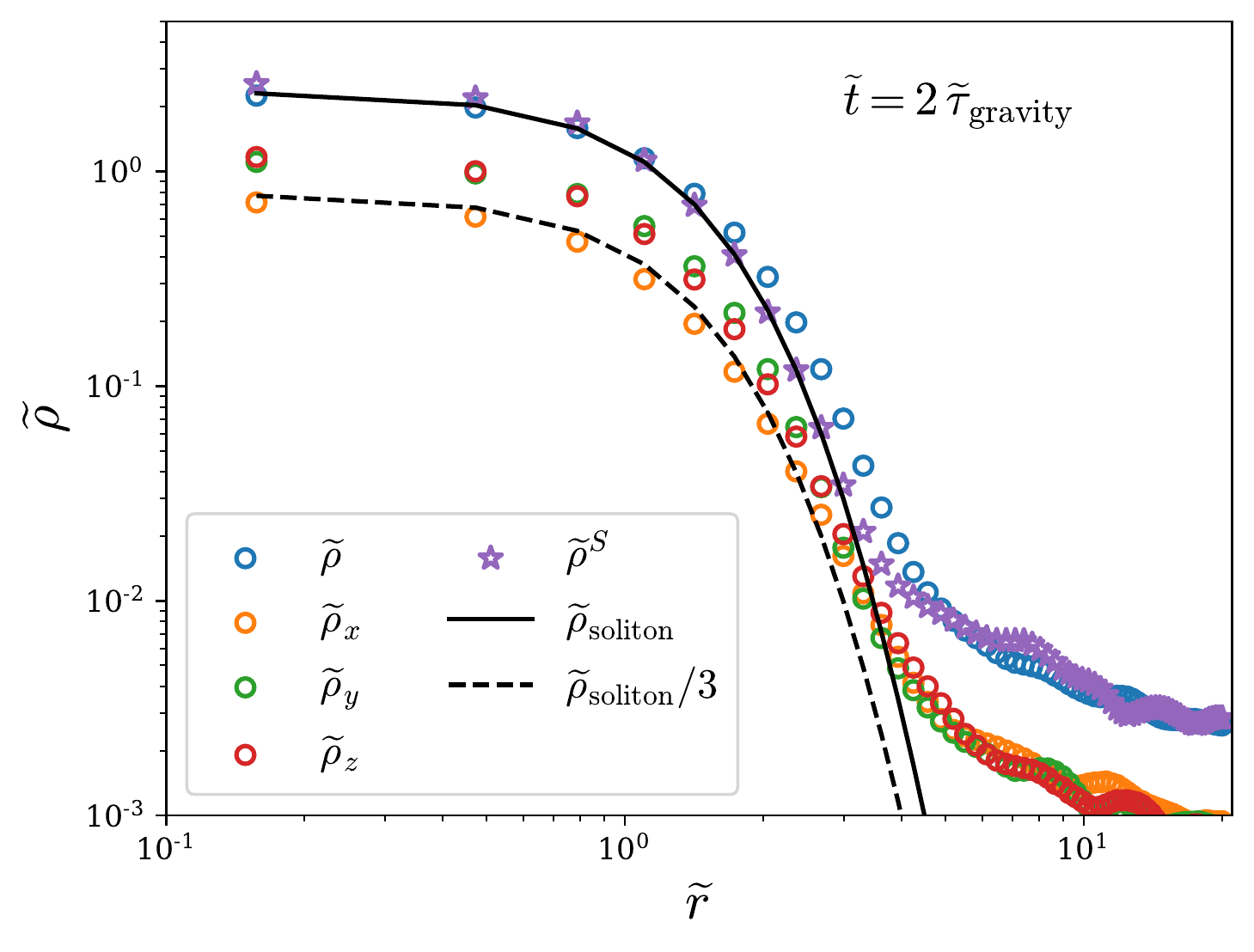}
\includegraphics[width=1.05\columnwidth]{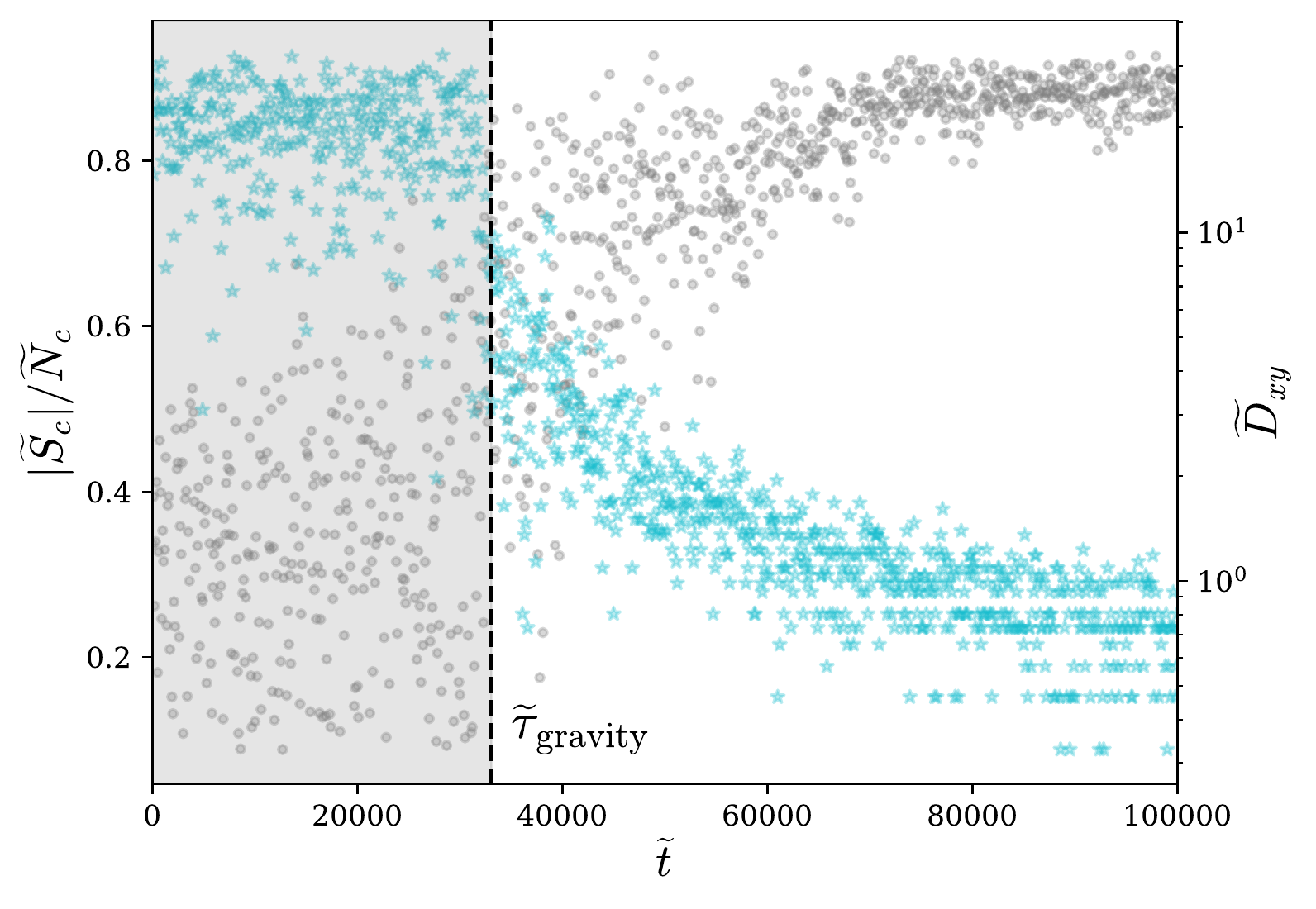}
\includegraphics[width=0.92\columnwidth]{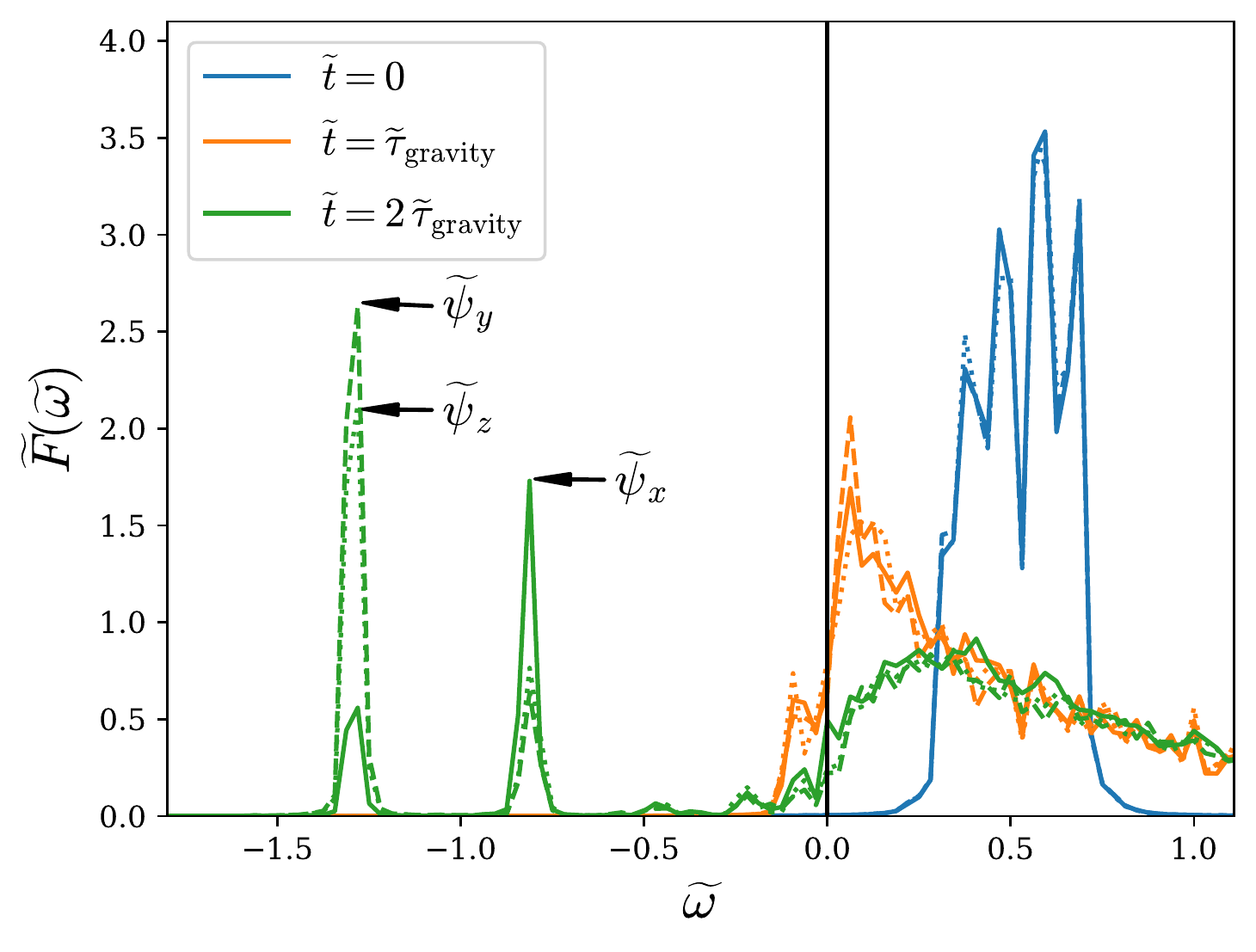}
\caption{Formation and evolution of the Proca star. The initial conditions are the same as Fig. \ref{fig:slice_N_84_L_42}. Upper left panel: Maximum of $|\bm{\Psi}|$, $|\psi_x|$, $|\psi_y|$ and $|\psi_z|$ as a function of time compared with the SDM case $|\psi^S|$. Upper right panel: radial density profile of the Proca star shown in the lower panel of Fig.~\ref{fig:slice_N_84_L_42} compared with the soliton profile (solid black curve) and the SDM case (purple stars). We also show the density profile of individual components of the vector field. Lower left panel: spin per particle in the core (left axis, grey circles), $|\bm{S}_c|/N_c$, and the offset between the positions where $|\psi_x|$ and $|\psi_y|$ are maximum, $\tilde{D}_{xy}$ (right axis, cyan stars). Note that at $t<\tau_{\rm gravity}$ (shaded region), the core radius is not well defined. At $t>\tau_{\rm gravity}$, the Proca star acquires a net spin angular momentum as it grows. Lower right panel: normalized energy spectrum at $t=0,\,\tau_{\rm gravity},\,2\,\tau_{\rm gravity}$.}
\label{fig:evo_N_84_L_42}
\end{figure*}

The maximum amplitude of the wave function (total and individual components) as a function of time is shown in the upper left panel of Fig.~\ref{fig:evo_N_84_L_42}. Inspired by the work by \cite{Levkov:2018kau}, we define the condensation time $\tau_{\rm gravity}$ as the time when the maximum density in the simulation box increases by a factor of $2$.\footnote{In practice, the maximum density fluctuates with time, so we first smooth the data before computing the condensation time using this definition.} At $t>\tau_{\rm gravity}$, we find that $|\bm{\Psi}|_{\rm max}$ grows linearly with time, similar to the findings in the SDM case \cite{Levkov:2018kau}. However, compared to SDM (purple circles), VDM (blue circles) condenses on a longer time scale. The condensation time is roughly $3$ times of that for SDM, confirming our prediction Eq.~(\ref{eq:tau_i}).

On the upper right panel of Fig.~\ref{fig:evo_N_84_L_42}, we show the radial density profile of the Proca star at $t=2\,\tau_{\rm gravity}$. As can be seen, the total density profile agrees with the soliton profile and the density profile of individual components is roughly $1/3$ of the total density profile. We also show the radial density profile of the scalar bose star (purple circles) that forms from initial conditions with the same mean density as the VDM simulation. Compared to the SDM case, the density profile of the Proca star slightly deviates from the soliton profile at large radii ($r>2 r_c$), indicating that the formed Proca star is slightly excited from the ground state. Ref.~\cite{Amin:2022pzv} studies the merger of VDM solitons and also shows that the final soliton is excited leading to a smoother transition from the solitonic core to the outskirt compared to the SDM case.

More specifically, if we look at the maximum density for individual components, the maximum density is not exactly at the same position in each case. In the lower left panel of Fig.~\ref{fig:evo_N_84_L_42}, we show the offset of position of the maximum density between the $x$ and $y$ components, $D_{xy}$ ($D_{xz}$ is similar, thus is not shown). As can be seen, after the formation of the Proca star ($t>\tau_{\rm gravity}$), the offset decreases with time, but does not decreases to zero at the end of our simulation. We also show the spin per particle in the core, $|\bm{S}|_c/N_c$ where $N_c$ is the number of nonrelativistic particles within the core radius defined in Eq.~\ref{eq:soliton_profile}. It is interesting that, even if the initial conditions do not contain net spin angular momentum, the formed Proca star does tends to acquire a net spin angular momentum. At $t\gg\tau_{\rm gravity}$, $|\bm{S}_c|/N_c \sim 0.8$. A similar phenomenon has been observed in simulations of the collision of multiple VDM solitons in \cite{Amin:2022pzv}.

Finally, in the lower right panel of Fig.~\ref{fig:evo_N_84_L_42}, we show the normalized energy spectrum of the system at different stages. For the definition of the energy spectrum, we take \cite{Levkov:2018kau}
\begin{equation}
F_i(\omega,t)=\int dt' d^3 x \,\psi_i^*(\bm{x},t)\psi_i(\bm{x},t+t')e^{i\omega t'-t'^2/\tau^2},
\label{eq:F_w}
\end{equation}
where $1/(m v_0^2) \ll \tau \ll \tau_{\rm gravity}$ is the width of the window function.~\footnote{In practice, we compute the energy spectrum by applying a Gaussian window function and doing FFTs along the time axis at each grid point, and then taking the spatial average.} We normalize the energy spectrum so that $\int \widetilde{F}_i(\widetilde{\omega}) d\widetilde{\omega}=1$. At the initial time, the spectrum peaks at $\widetilde{\omega} \sim 0.5$ due to the choice of initial momentum distribution, i.e. $\omega\sim m v_0^2/2$. As condensation happens at $\tau_{\rm gravity}$, a peak appears at a negative frequency indicating the formation of a bound state. At $2\,\tau_{\rm gravity}$, the peak moves toward more negative values and the amplitude becomes higher. However, we find that the frequencies of the highest peaks in the region $\widetilde{\omega}<0$ are different for different components (arrows in the lower right panel of Fig.~\ref{fig:evo_N_84_L_42}). As mentioned earlier, this is due to the excited modes in the Proca star. But we do see that at the frequency corresponding to the highest peak of one component, e.g. $\widetilde{\omega}\sim-0.81$, there exist secondary peaks in the spectrum of other components, too.

\begin{figure}[htbp]
\includegraphics[width=\columnwidth]{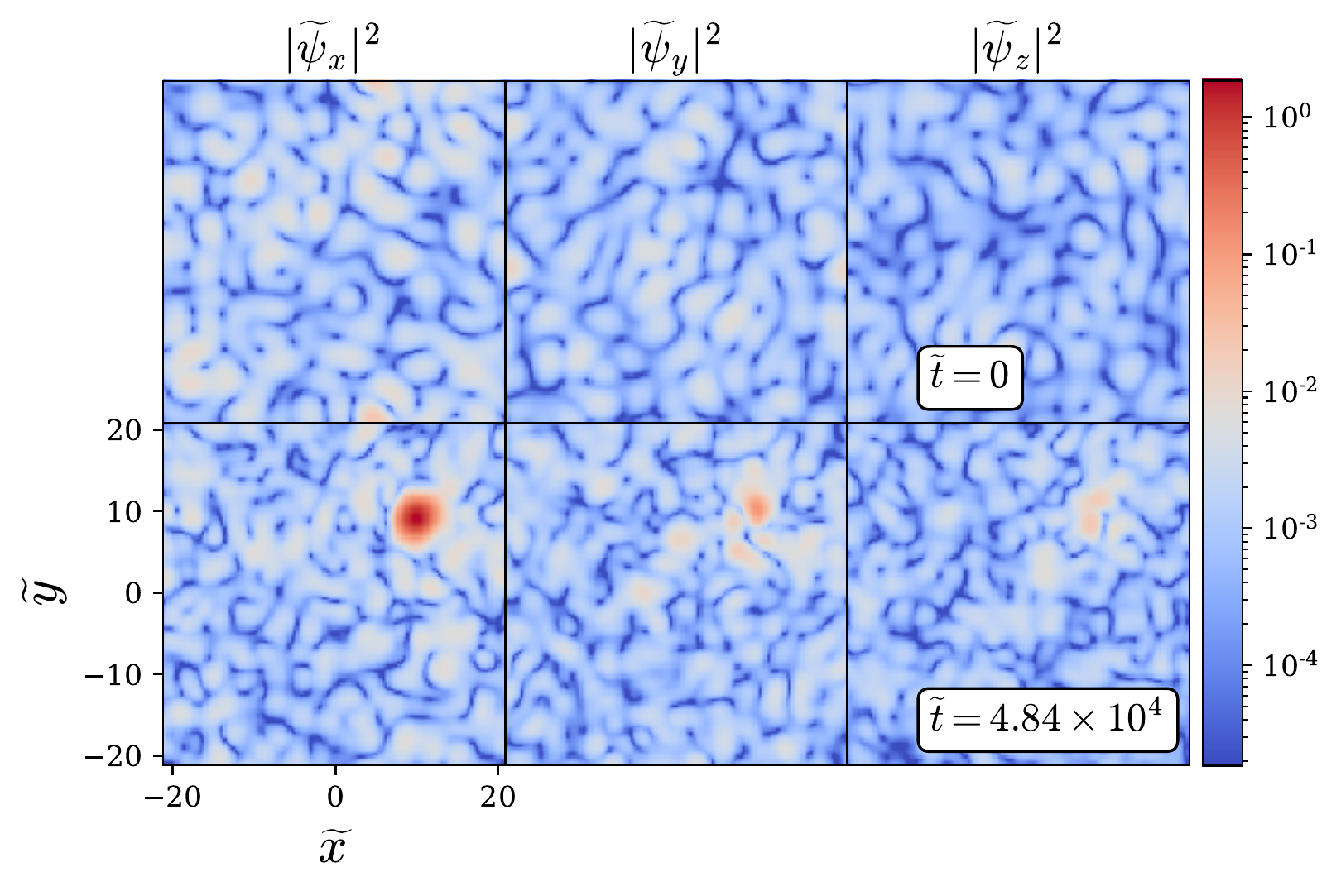}
\caption{Formation of a Proca star from homogeneous and isotropic initial conditions with $\widetilde{N}_1=113.1$, $\widetilde{N}_2=80.4$, $\widetilde{N}_3=57.8$, and $\widetilde{L}=42$. Upper panel: slice density field for each vector component at the initial time. Lower panel: slice density field through the center of the Proca star at $\widetilde{t}=2\,\widetilde{\tau}_{\rm gravity}$.}
\label{fig:slice_N_113_80_58_L_42}
\end{figure}

For initial conditions with mass equally distributed among different components, we expect all VDM components to condense on similar time scales. However, things will be different if one component has higher mean density than other components. To explore this, we also run simulations with unequal mass distribution among different components. One example is shown in Fig.~\ref{fig:slice_N_113_80_58_L_42} where we take $\widetilde{N}_1=113.1$, $\widetilde{N}_2=80.4$, $\widetilde{N}_3=57.8$, and $\widetilde{L}=42$. Again, a Proca star is found to form in the simulation box. But the mass of Proca star mainly comes the component that has the highest initial density, i.e. $\psi_x$. At the position of the Proca star, overdensities also appear in the other two components, $\psi_y$ and $\psi_z$, but the amplitude of the overdensity is much smaller.

\begin{figure*}[t]
\includegraphics[width=\columnwidth]{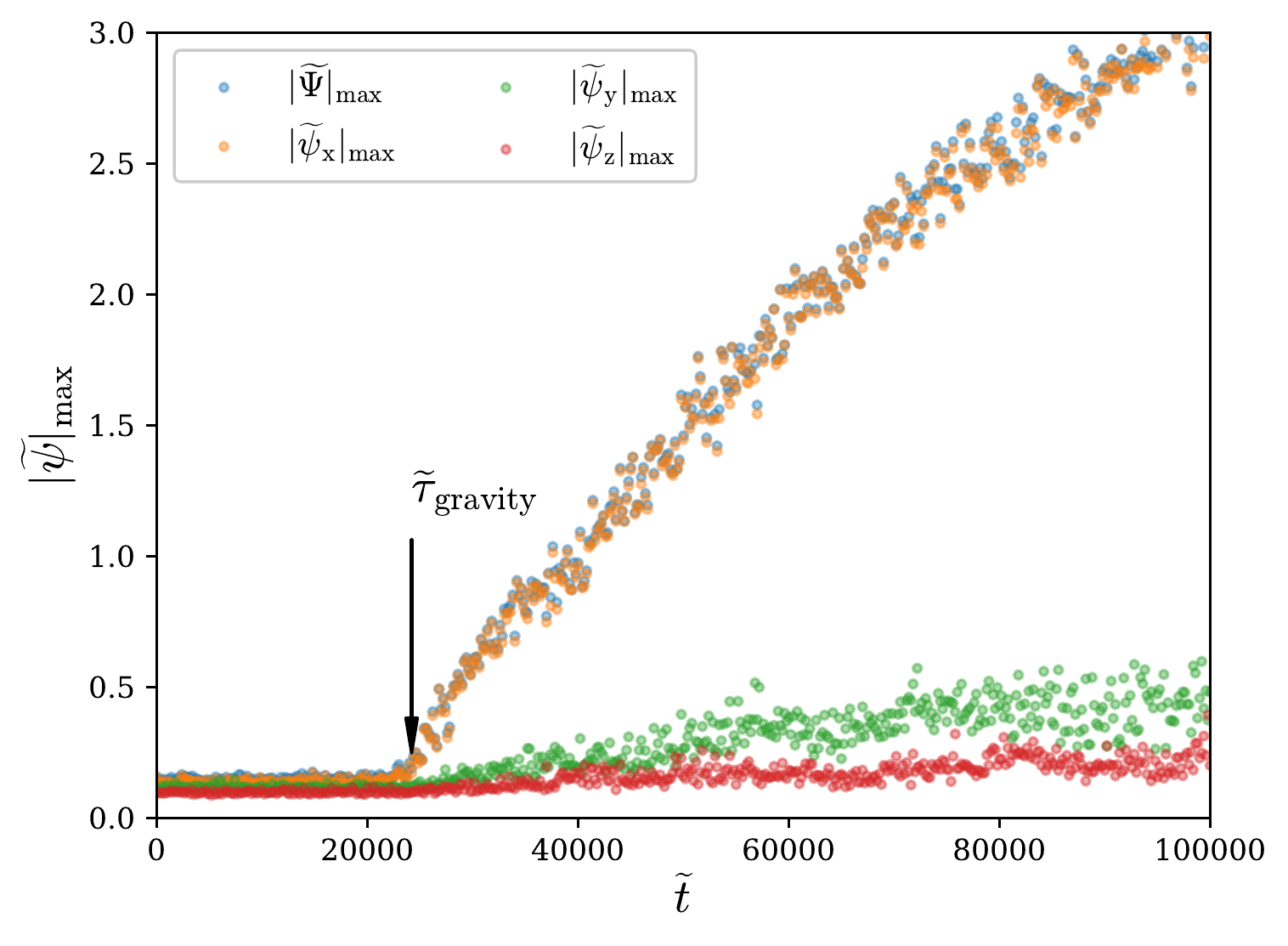}
\includegraphics[width=0.95\columnwidth]{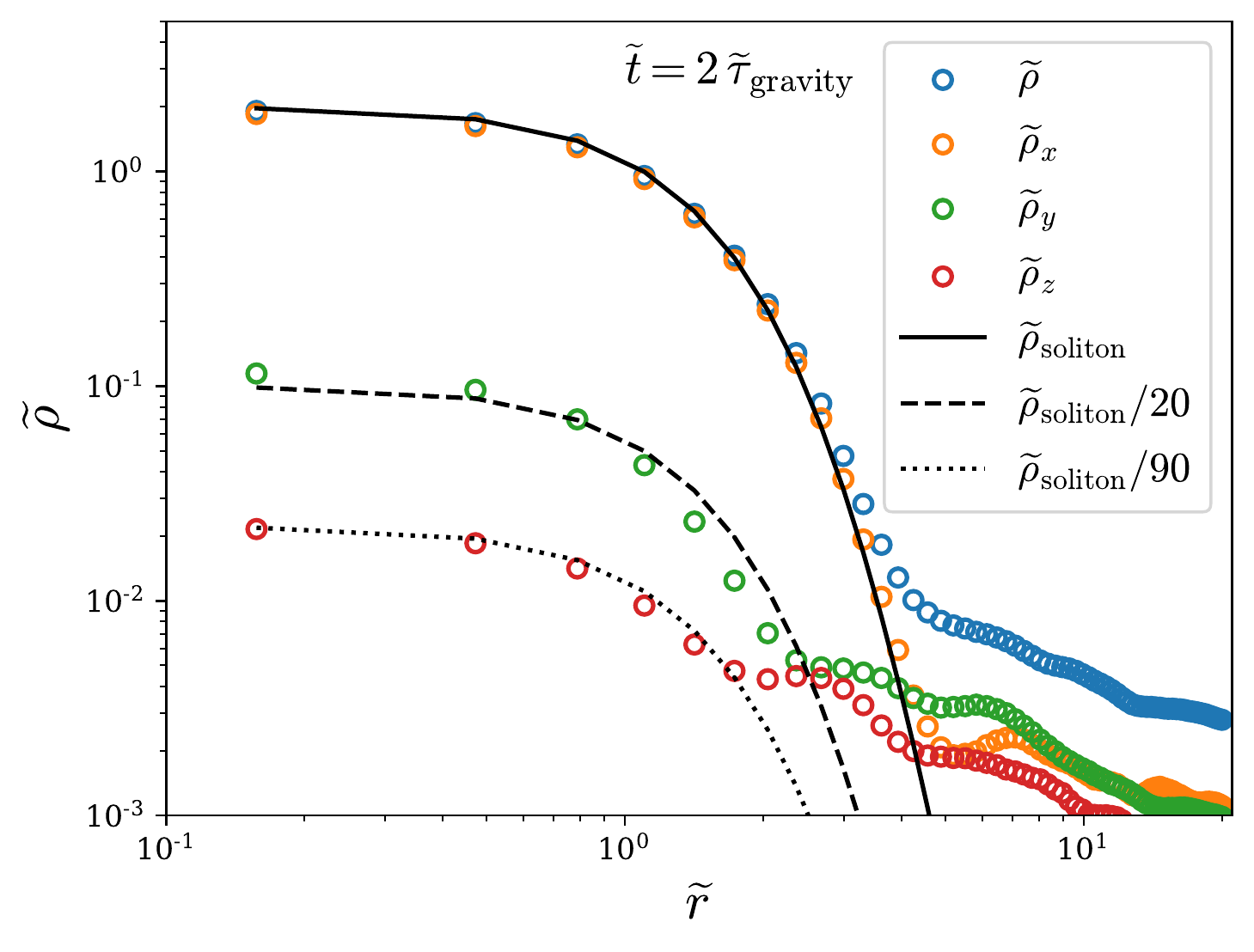}
\caption{Formation and evolution of the Proca star. The initial conditions are the same as that in Fig.~\ref{fig:slice_N_113_80_58_L_42}. Left panel: Maximum of $|\bm{\Psi}|$, $|\psi_x|$, $|\psi_y|$ and $|\psi_z|$ as a function of time. Right panel: radial density profile of the Proca star compared with soliton profiles at $t=2\,\tau_{\rm gravity}$.}
\label{fig:evo_N_113_80_58_L_42}
\end{figure*}

In the left panel of Fig.~\ref{fig:evo_N_113_80_58_L_42}, we show the evolution of maximum amplitude of the wave function from the above simulation. We find that although the condensation time scales of $\psi_y$ and $\psi_z$ are expected to be longer from the prediction of Eq.~(\ref{eq:tau_i}), $|\psi_y|_{\rm max}$ and $|\psi_z|_{\rm max}$ start to grow at the same time as $|\psi_x|_{\rm max}$. This is likely due to the infall of VDM particles into the gravitational potential of the newly formed Proca star. Thus we propose that the condensation time scale is determined by the component with the highest initial density, i.e.
\begin{equation}
\tau_{\rm gravity} = \min\{\tau_{i,gravity}\}.
\label{eq:tau_t}
\end{equation}

The radial density profiles of the Proca star is shown in the right panel of Fig.~\ref{fig:evo_N_113_80_58_L_42}. As mentioned above, the mass is mainly contributed by the $x$ component, whose profile agrees well with the soliton profile. The amplitude of the density profiles (green and red circles) for the $y$ and $z$ components are much lower, but their shape is close to the soliton profile (dashed and dotted curves) in the central region.

\begin{figure}[htbp]
\includegraphics[width=\columnwidth]{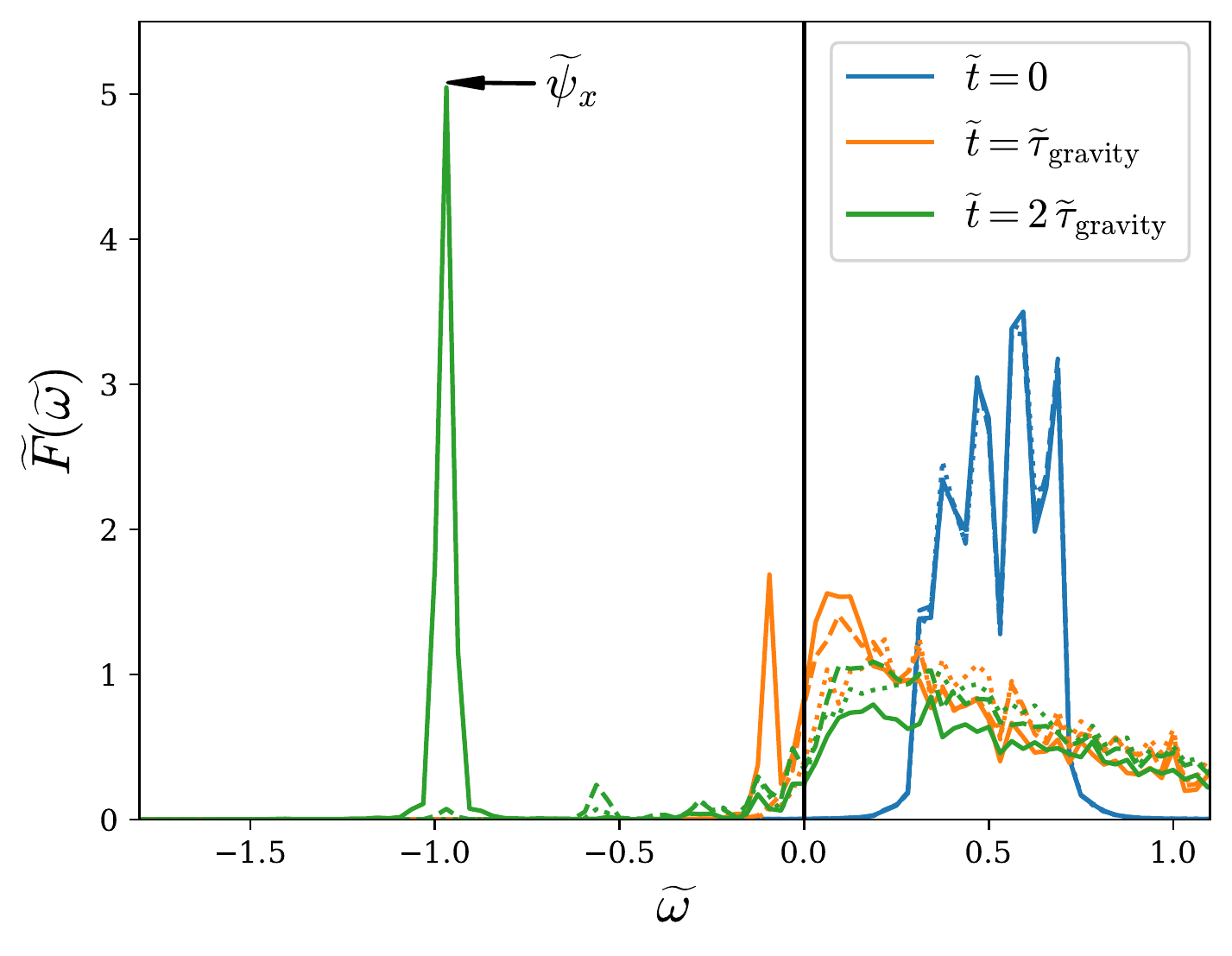}
\caption{Normalized energy spectrum at different times for the same initial conditions as in Fig.~\ref{fig:slice_N_113_80_58_L_42}.}
\label{fig:spectrum_N_113_80_58_L_42}
\end{figure}

For this case, we also analyze the energy spectrum before and after the formation of the Proca star, as shown in Fig.~\ref{fig:spectrum_N_113_80_58_L_42}. Similar to the previous case with equally distributed mass among the components, at $t=\tau_{\rm gravity}$, a peak appears at a negative frequency. However, at $t=2\,\tau{\rm gravity}$, we find only one prominent peak in the region $\widetilde{\omega} < 0$ which appears at $\widetilde{\omega}\sim -0.97$ in the spectrum of the $x$ component. At the same frequency, we again see enhancements in the spectra of the other two components, although the effect is very tiny. Besides, several peaks with small amplitudes are seen between $-0.97 < \widetilde{\omega} < 0$. They correspond to the eigenenergies of excited modes in the gravitational potential of the Proca star. Note that the potential is dominated by the $x$ component.

\subsection{Partially correlated initial conditions}\label{subsec:partial_corr}
In the previous subsection, we consider uncorrelated initial conditions. But in the real Universe, correlation may grow during structure formation leading to partially correlated configurations in the halo center where the Proca star forms. Therefore, in this subsection we will look at the case with partially correlated initial conditions. To quantify the correlation between different components, we define a covariance matrix as
\begin{equation}
C_{ij}=\langle \psi^*_i \psi_j \rangle,
\label{eq:C_ij}
\end{equation}
where $\langle.\rangle$ denotes ensemble average which can be replaced by spatial average according to the Ergodic theory. For uncorrelated initial conditions, the covariance matrix is diagonal with the diagonal part the mean number density of each component. For correlated initial conditions, $\{C_{ij}\}$ is a general Hermitian matrix with real eigenvalues. We can always find a unitary transformation $U$ that diagonalizes the covariance matrix: $D=U^{-1} C U$. In other words, we can choose the normalized eigenvectors of $\{C_{ij}\}$ as a new basis to make different components of the vector field uncorrelated. Under the new basis, $\bm{\Psi}'=U^{-1}\bm{\Psi}$ and the mean number densities of individual components of $\bm{\Psi}'$ are the eigenvalues of $\{C_{ij}\}$. As discussed in Sec. \ref{sec:symmetry}, the condensation time is unaffected by unitary transformations. So we can use the eigenvalues of $\{C_{ij}\}$ and the prediction for uncorrelated initial conditions, Eqs. (\ref{eq:tau_i}) and (\ref{eq:tau_t}), to estimate the condensation time for correlated initial conditions.

To check the above conclusion, we create initially correlated vector fields. We first generate two independent Gaussian random fields with $\delta$ momentum distribution in the same way as we do in the previous subsection. Then we take
\begin{eqnarray}
\psi_x &=& \psi_1,
\label{eq:corr_ini_x}\\
\psi_y &=& C \psi_1 + \sqrt{1-C^2} \psi_2.
\label{eq:corr_ini_y}
\end{eqnarray}
Without loss of generality, we set $n_1=n_2$ with $n_1=\langle\psi_1^*\psi_1\rangle$ and $n_2=\langle\psi_2^*\psi_2\rangle$, and $\psi_z=0$. With such initial configurations, $\psi_x$ and $\psi_y$ are initially correlated. The correlation is controlled by the coefficient $C$.

We find that if we rotate the field given by Eqs.~(\ref{eq:corr_ini_x}) and (\ref{eq:corr_ini_y}) with respect to the $z$ axis by an angle of $\pi/4$, the new field becomes
\begin{eqnarray}
\psi_x' &=& \frac{\sqrt{2}}{2}\left( 1-C \right) \psi_1-\frac{\sqrt{2}}{2}\sqrt{1-C^2}\psi_2, 
\label{eq:psi_x_rot}\\
\psi_y' &=& \frac{\sqrt{2}}{2}\left( 1+C \right) \psi_1+\frac{\sqrt{2}}{2}\sqrt{1-C^2}\psi_2.
\label{eq:psi_y_rot}
\end{eqnarray}
Considering that $\langle \psi_1^* \psi_2 \rangle=0$ \footnote{$\psi_1$ and $\psi_2$ are independent Gaussian random fields.} and $n_1=n_2$, it is easy to check that $C_{xy}' = 0$, i.e. the different components of the new vector field are uncorrelated. The mean number density of the components of the new vector field are given by
\begin{eqnarray}
n_x' &=& (1-C) n_1
\label{eq:nx_rot}\\
n_y' &=& (1+C) n_2,
\label{eq:ny_rot}\\
n_z' &=& 0.
\label{eq:nz_rot}
\end{eqnarray}
Since the rotation is a special type of unitary transformation, it does not affect the condensation time. So we expect the original partially correlated vector field to condense on the same time scale as the uncorrelated initial conditions with mean number densities given by Eqs. (\ref{eq:nx_rot}), (\ref{eq:ny_rot}) and (\ref{eq:nz_rot}). To check this, we run two sets of simulations: one set with correlated initial conditions given by Eqs. (\ref{eq:corr_ini_x}) and (\ref{eq:corr_ini_y}); the other set with uncorrelated initial conditions but with mean number density given by the above equations. \footnote{Note that these initial conditions are generated from independent random seeds and not by rotation.} For the simulation parameters, we take $\widetilde{N}_1=\widetilde{N}_2=125.7$ and $\widetilde{L}=62.5$.

\begin{figure}[htbp]
\includegraphics[width=\columnwidth]{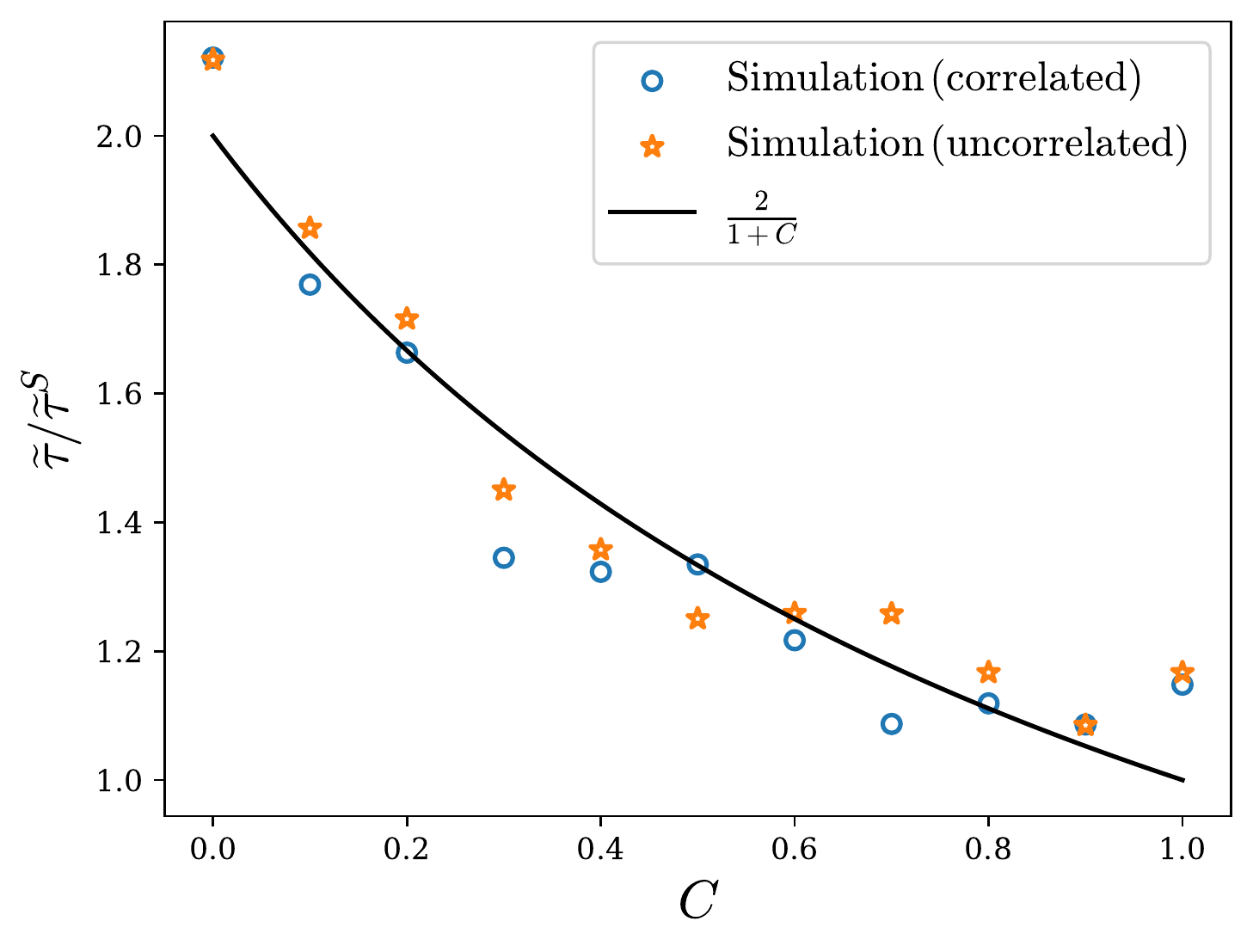}
\caption{Condensation time for initial conditions with different correlations from simulations (open circles) compared to the uncorrelated initial conditions with mean number densities given by Eqs. (\ref{eq:nx_rot}) (\ref{eq:ny_rot}), and (\ref{eq:nz_rot}) (open stars). The black solid line shows the prediction from Eqs. (\ref{eq:tau_i}) and (\ref{eq:tau_t}). Here we plot the condensation time for VDM in units of the condensation time of the scalar field with the same mean total number density.}
\label{fig:tau_corr}
\end{figure}

The condensation times for different values of $C$ from the above two sets of simulations are shown in Fig.~\ref{fig:tau_corr}. As can been seen for a fixed value of $C$, the condensation time scales for two sets of simulations are similar as expected and both agree with the analytic formula Eqs. (\ref{eq:tau_i}) and (\ref{eq:tau_t}) (solid black curve) if we plug in $n_i'$ instead of $n_i$.

\section{Relaxation time of Bose-Einstein condensation by gravity}\label{sec:relax_time}

\begin{figure}[htbp]
\centering
\includegraphics[width=\columnwidth]{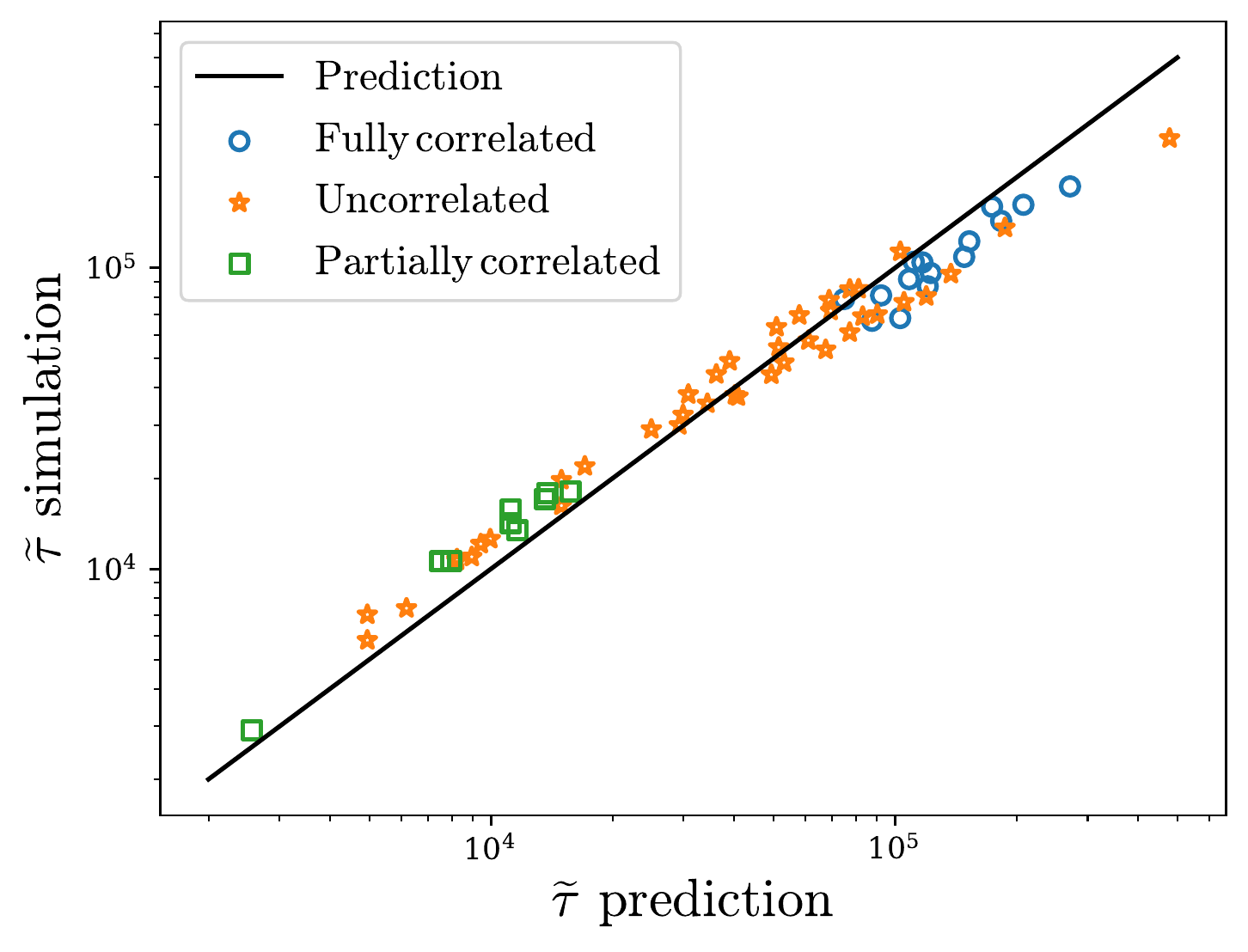}
\caption{Condensation time of VDM with gravity for different initial conditions. The prediction is computed using Eqs. (\ref{eq:tau_i}) and (\ref{eq:tau_t}) and take into account the possible correlation between different vector components as discussed in Sec. \ref{subsec:partial_corr}.}
\label{fig:vdm_tau_fit}
\end{figure}

Following the discussions in the previous section, we have created a number of different initial conditions with different box sizes, mean number density, and with and without correlations and have run numerical simulations to measure the condensation time scale in each simulation. The results are summarized in Fig.~\ref{fig:vdm_tau_fit}. We find good agreement between the simulation results (colored markers) and the analytic formula (black solid curve) we proposed in the previous section, i.e. using Eqs. (\ref{eq:tau_i}) and (\ref{eq:tau_t}) and taking into account the possible correlation between different vector components as described in Sec. \ref{subsec:partial_corr}. The coefficient in Eqs. (\ref{eq:tau_i}) is taken to be $0.7$, similar to the value for SDM.

\section{Discussion and Conclusions}\label{sec:conclusion}

Using numerical simulations of the Schr\"{o}dinger-Poisson equations for VDM in the non-relativistic limit, we studied the condensation of VDM due to long-range gravitational interaction. We find that Proca stars can form dynamically from a homogeneous and isotropic initial distribution of VDM. The Proca star has a radial density profile similar to the scalar Bose star, but is perturbed by excited modes (Fig.~\ref{fig:evo_N_84_L_42}).

Compared to SDM, VDM contains extra degrees of freedom. The condensation time scale depends on the correlation between its different components. For fully correlated initial conditions, e.g. extremely polarized configurations, VDM behaves exactly like SDM, i.e. condensing with the same rate.

For uncorrelated initial conditions, the condensation time is longer in the VDM case compared to the SDM case with the same mean number density. If the mass is equally distributed among different components in the initial conditions, we find the condensation is $3$ times longer than the corresponding SDM case (Fig.~\ref{fig:evo_N_84_L_42}). The amplitude of the Proca star wave function grows linearly with time similar to the SDM case. However, as the Proca star grows, it acquires a net spin angular momentum even if the initial conditions do not contain any net spin angular momentum. The average spin per particle in the core approaches $\sim\hbar$ at $t \gg \tau_{\rm gravity}$. After the formation of the Proca star, we find that it is perturbed by excited modes. The effect is more significant than SDM, leading to a small deviation of the radial density profile from the soliton profile beyond $2\,r_c$. When the Proca star forms in the simulation box, we notice a peak at negative frequency appears in the energy spectrum similar to SDM. However, as the Proca star grows, the highest peak in the spectrum of different components may shift to different frequencies, again indicating the important influence of the excited modes. If the mass is unequally distributed among different components, we find the component with the highest mean number density dominates the condensation, while the other components start to condense roughly at the same time, but at a much slower rate.

Finally, for partially correlated initial conditions, we find that we can transform the correlated configuration to an uncorrelated one by a unitary transformation. The unitary transformation does not affect the condensation time. Thus we can use our results for uncorrelated initial conditions to estimate the condensation time scale for correlated ones. The mean number density of different vector component after the unitary transformation are the eigenvalues of the covariance matrix (Eq.~\ref{eq:C_ij}).

Understanding the condensation of VDM will be useful for estimating how likely Proca stars are to have formed in the Universe through the kinetic regime and shed light on whether we can distinguish VDM from SDM phenomenologically. In the current work, we neglect the possible self-interactions which may contain a spin-spin interaction term. A spin-spin interaction would mix up different components of the vector field, e.g. the total number of particles in each individual component is no longer conserved, leading to rich phenomena that do not occur for SDM \cite{Amin:2022pzv}. We will study the effect of self-interactions on condensation and formation of Proca stars in a future work.

\textbf{Note}. As this work was put online, another work by Jain et al. \cite{Jain:2023ojg} also appeared on arXiv. Ref.~\cite{Jain:2023ojg} derived the wave-kinetic Boltzmann equation for multicomponent bosonic dark matter (of which VDM is a subset) and obtained the condensation time under the eikonal approximation. For models with three components of the same particle mass, the results of Ref.~\cite{Jain:2023ojg} correspond to VDM and their results are in agreement with ours for uncorrelated initial conditions. In this work, we considered in addition partially correlated initial conditions, demonstrating the importance of such correlations and an understanding of them based on $U(N)$ transformations.

\section*{Acknowledgements}
DJEM is supported by an Ernest Rutherford Fellowship from the Science and Technologies Facilities Council (ST/T004037/1).
JC is supported by Fundamental Research Funds for the Central Universities(5330500858).
XD thanks Yang Bai for beneficial discussions. JC and XD acknowledge helpful discussions with Mudit Jain after the first submission of this work to arXiv.

This research made use of computational resources at the School of Physical Science and Technology, Southwest University. Part of the computations presented here were conducted at the Resnick High Performance Computing Center through Carnegie's partnership in the Resnick Sustainability Institute at the California Institute of Technology.

\bibliography{reference}

\end{document}